\documentclass[aps,pra,twocolumn]{revtex4}
\usepackage{}
\usepackage{amssymb}
\usepackage{color}
\usepackage{amsmath}
\usepackage{graphicx}
\usepackage{subfigure}
\usepackage{txfonts}
\usepackage{color}
\usepackage{ulem}

\newcommand{\fisicarm}{Dipartimento di Fisica, Sapienza Universit\`{a} di Roma, Piazzale Aldo Moro, 5, I-00185 Roma, Italy}
\newcommand{\fisicami}{Istituto di Fotonica e Nanotecnologie - CNR and Dipartimento di Fisica - Politecnico di Milano, Piazza Leonardo da Vinci, 32, I-20133 Milano, Italy}
\newcommand{\pisasns}{NEST, Scuola Normale Superiore and Istituto di Nanoscienze - CNR, I-56126 Pisa, Italy}

\newcommand{\out}{\mathrm{out}}
\newcommand{\inp}{\mathrm{in}}

\begin{document}

\title{All-optical non-Markovian stroboscopic quantum simulator}
\author{Jiasen Jin, Vittorio Giovannetti, and Rosario Fazio}
\affiliation{\pisasns}

\author{Fabio Sciarrino and Paolo Mataloni}
%\affiliation{\ifn}
\affiliation{\fisicarm}

\author{Andrea Crespi and Roberto Osellame}
%\affiliation{\ifn}
\affiliation{\fisicami}
\begin{abstract}
An all-optical scheme for simulating non-Markovian evolution of a quantum system is proposed. It uses only linear optics elements and by controlling the system parameters allows one to control the presence or absence of information backflow from the environment. A sufficient and necessary condition for the non-Markovianity of our channel based on Gaussian inputs is proved. Various criteria for detecting non-Markovianity are also investigated by checking the dynamical evolution of the channel.
\end{abstract}
\maketitle

\section{Introduction}

In the absence of memory effects in the system environment,
the dynamics of open quantum systems is typically characterized in terms of master equations in the so-called Lindblad form~\cite{OPEN}. This corresponds to having a continuous Markovian process which, while being an extremely useful approximation, does not necessarily hold at all timescales. In recent years a huge effort has been put forward in order to characterize non-Markovian behaviors in the dynamics of quantum systems~\cite{VAC,PII,NORI,CHRU,RIVAS,SAB,BR,BR1,rS1,rS2,hou,BUSCEMI,WOLF}.
Understanding these phenomena is a fundamental question with potential applications in the engineering of reservoirs for quantum computation and quantum information processing \cite{VERS,VASILE,MATS,TEL,COM}.
Most of the works proposed so far have focused on developing tools and criteria that would permit one to certify the presence of a non-Markovian behavior~\cite{RIVAS,SAB,BR,BR1,rS1,rS2,hou,BUSCEMI}, relying on certain  mathematical properties of this type of dynamical evolution~\cite{MAT1,MAT2,MAT3,WOLF}. Recently a couple of works in simulating and testing non-Markovian dynamics in optical systems have been reported \cite{Chiuri,Orieux,Zou}.

An alternative approach to the study of non-Markovianity has been proposed in a series of papers based on collisional models~\cite{COLL1,COLL2,COLL3,COLLnonM1,COLLnonM2,COLLnonM3,COLLnonM4,BERN}.
Here the continuous time evolution is replaced by a stroboscopic process formed by  a series of discrete steps where the system of interest couples with different components of a many-body quantum system that simulates an extended environment.
Despite being extremely simplified  such models allow one to capture the main effects which are responsible for the arising of the non-Markovian character in the dynamics, namely, the correlations in the environment input state, its internal dynamics, and the interaction between the latter and the system. In particular Refs.~\cite{COLL1,COLL2} presented a general scheme where the non-Markovianity is  simulated by allowing collisional interactions  also among the various components of the extended many-body quantum environment. By properly  tuning such intra-environment collisions one can then pass from a purely Markovian dynamics to a strong non-Markovian regime where the system environment reduces to a single, finite dimensional quantum system which coherently exchange information with the system of interest via conventional Rabi oscillations. As a matter of fact such a theoretical scheme can be perceived as  a quantum simulator that enables one to reproduce (in a stroboscopic fashion) the (not necessarily Markovian) complex dynamics of an open quantum system.

The aim of this work is to propose an implementation of such a theoretical model using linear quantum optics. Specifically in our scheme the system of interest as well as the environmental degree of freedom are all described in terms of frequency-degenerate quantum optical modes while their interactions are represented as a complex scattering process mediated by a series of properly arranged beam splitters (BSs).
The overall arrangement is very much similar to the one used to test interaction-free measurement \cite{KWIAT}. The main difference is that in our case thermal sources have to be injected into the auxiliary ports of the setup. In this scenario we explicitly solve the stroboscopic dynamics of the system and study the arising of non-Markovianity as a function of the setup parameters (e.g. the transmissivities of the BSs, the temperature of the environmental modes, the relative phase accumulated by the signals between two consecutive collisions). Hence we test the non-Markovian witnesses of Refs. \cite{RIVAS,SAB,BR,BR1,rS1,rS2} and compare their efficiencies.

The paper is organized as follows. We start in Sec.~\ref{scheme} by describing the details of our proposal. In Sec.~\ref{characterization} we present a sufficient and necessary characterization of the non-Markovianity of stroboscopic evolution. In Sec. \ref{NONM} instead we analyze how different non-Markovianity witnesses perform in our model. The conclusions are drawn finally.

\section{Scheme}\label{scheme}
Our scheme is based on the theoretical model discussed in Refs.~\cite{COLL1,COLL2} where the non-Markovian dynamical evolution of a quantum system $S$ is stroboscopically represented via a series of (coherent or incoherent) collisions with a collection of environmental ancillary quantum systems. Accordingly dissipation and decoherence are induced at each collision by allowing  part of the ``information" contained in $S$ to leak into the environment. However, at variance with standard memoryless collisional models~\cite{COLLnonM1,COLLnonM2,COLLnonM3,COLLnonM4}, in the scheme of Refs.~\cite{COLL1,COLL2} the bath is endowed with an effective memory by introducing extra inter-ancillary collisions between two consecutive system-ancilla interactions. By controlling the intensity of such inter-ancillary collisional events, the model allows one to interpolate between a fully Markovian dynamics (where no back-flow of information from the environment to $S$ is allowed) and the continuous interaction of the system with a single ancilla, i.e., a strongly non-Markovian process.

\begin{figure}
  \includegraphics[width = 1\columnwidth]{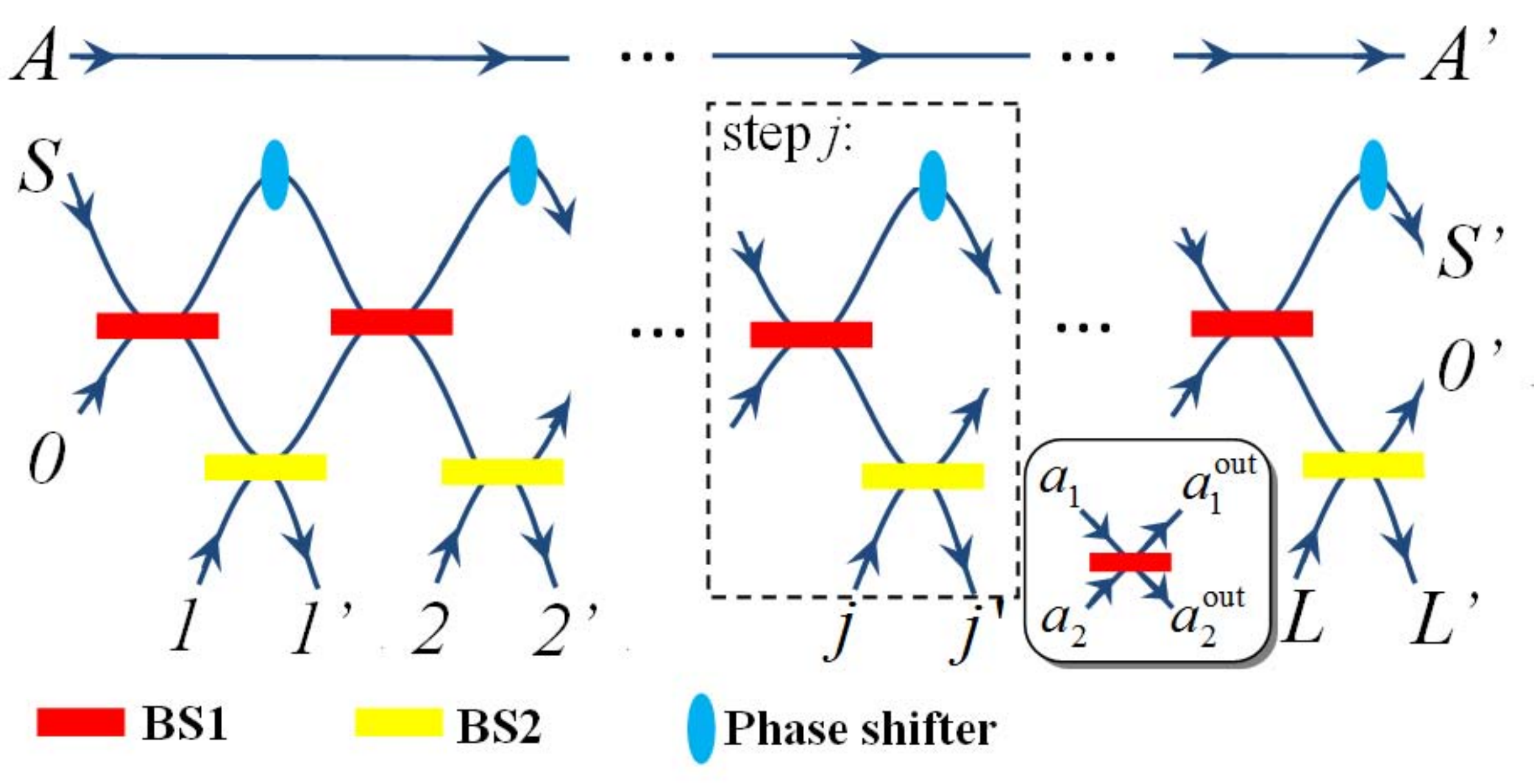}
  \caption{(Color online). Schematic of the setup. The BSs in red represent collisions between the system of interest (mode $S$) and one of the environments (modes $0,1,2,...,L$). The BSs in yellow instead describe intra-environment collisions: their reflectivities $r_2$ define the back-flow of information into the system (for $r_2=0$ the model becomes strictly Markovian).
  An auxiliary mode $A$, which is entangled with mode $S$, is introduced to detect the non-Markovianity of the stroboscopic evolution as discussed in Sec. \ref{NONM}. The solid-line box is a description of the interaction of two input modes at the BS. The dashed-line box denotes the building block of our stroboscopic model. The prime denotes the output mode.} \label{fig1}
\end{figure}

The collision model can be implemented in an all-optical system where the system and the ancillary environments are represented by independent (frequency-degenerate) modes propagating along different optical paths, while the collisional events among them are simulated by means of BSs. We recall that a BS transfers the input modes, say $a_1$ and $a_2$, in the following way:
\begin{equation}
\left(
  \begin{array}{c}
    a_1^{\mathrm{out}} \\
    a_2^{\mathrm{out}} \\
  \end{array}
\right)=\left(
          \begin{array}{cc}
            r & t \\
            t & -r \\
          \end{array}
        \right)\left(
                 \begin{array}{c}
                   a_1 \\
                   a_2 \\
                 \end{array}
               \right),
\end{equation}
where $r$ and $t$ are the reflectivity and transmissivity of the BS satisfying $r^2+t^2=1$ (see the solid-line box of Fig. \ref{fig1}). The mixture of two input modes on the BS is an effective simulation of the information exchange.

The setup is composed by an array of BSs described in Fig.~\ref{fig1}. An optical incoming signal (representing the system mode) prepared in state $\rho_S$ enters from the port $S$ and interacts with the $0$th environmental mode (entering from port $0$), prepared in a given state $\rho_0$ which will be specified later, at the first BS1. After this interaction, a part of the information will be reflected back to the system $S$, while the other part is transmitted into the environment. However, as it is clear from Fig. \ref{fig1}, the lost information from $S$ has a chance of reentering in the system due to the presence of the BS2 associated with the first environmental mode (entering the system in a given state $\rho_1$ from port $1$) and of the second BS1. In between two neighboring BS1s, there is a phase shifter which introduces a phase difference $\phi$ between the system and the memory. As a consequence, we can obtain the dynamical evolution of $\rho_S$ by concatenating such processes as a building block and introducing more environmental modes.

To solve the stroboscopic evolution of $\rho_S$ one needs to compute the scattering matrix ${\cal S}$ of the channel. For a channel with environmental modes labeled by $0,1,...,L$, ${\cal S}$ is an ($L+2$)-dimensional matrix. It can be represented as ${\cal S}(L)=\prod_{j=1}^{L}{{\cal S}_j}$ where ${\cal S}_j$ are the following:
\begin{equation}
{\cal S}_{j\geq1}=\left(
      \begin{array}{ccccc}
        r_1e^{i\phi} & t_1e^{i\phi} & 0 & 0 & 0 \\
        t_1r_2 & -r_1r_2 & 0 & t_2 & 0 \\
        0 & 0 & I_{j-1} & 0 & 0 \\
        t_1t_2 & -r_1t_2 & 0 & -r_2 & 0 \\
        0 & 0 & 0 & 0 & I_{L-j} \\
      \end{array}
    \right),
\end{equation}
with $I_{m}$ being an $m\times m$ identity and $r_1 (t_1)$ and $r_2 (t_2)$ being the reflectivities (transmissivities) of BS1 and BS2, respectively.
The parameter $r_1$ measures the intensity of the interaction between the system and the environment, while $r_2$ is the memory parameter of the channel as it measures the back-flow of information. In particular, for $r_2=0$ no back flow is allowed and the channel becomes perfectly Markovian.

The dynamical evolution of the system plus the environment can then be solved using the characteristic function formalism~\cite{WM}. The multi mode (symmetrically ordered) characteristic function of the input modes of the joint system is given by
\begin{equation}
\chi_{J}^{\inp}(\vec{\nu})= \mbox{Tr} [ D(\vec{\nu}) \rho_{J}^{\inp}],
\end{equation}
where $\rho_{J}^{\inp}$ is the joint state of all input modes and $\vec{\nu}=(\nu_S,\nu_0,\nu_1,...,\nu_L)$ is a complex vector (the subscripts $S$ and $i=0,1,...,L$ denoting respectively the system and the $i$th environmental mode). Accordingly, the joint characteristic function of the output modes is given by
\begin{equation}
\chi_{J}^{\out,L}(\vec{\nu})
=\chi_{J}^{\inp}[\mathcal{S}^{-1}(L)\vec{\nu}],
\end{equation}
from which  the characteristic function $\chi_{{S}}^{\out,L}(\nu_S)$ associated with the output of the system mode $S$ can be obtained
by simply setting $\vec{\nu}=(\nu_S,0,...,0)$, i.e.,
\begin{eqnarray}
 \chi_{{S}}^{\out,L}(\nu_S)= \chi_{J}^{\out,L}(\nu_S,0,...,0)\;.
 \end{eqnarray}

Assuming then that the environmental modes are all initialized in the same thermal state, the above expression yields the following input-output relation for the system mode $S$,
\begin{eqnarray}\label{EQ0}
\chi^{\inp}_S(\nu_S)
&\mapsto&
\chi^{\out,L}_S(\nu_S)\cr\cr&=&\exp{[(n_T+\frac{1}{2})(|c_L|^2-1)
|\nu_S|^2}]\chi^{\inp}(c_L^*\nu_S),
\end{eqnarray}
where $\chi^{\inp}(\nu_S)$ is the characteristic function of input state $\rho^{\inp}_S$ of $S$,
 $n_T$ is the thermal photon number of the environmental modes, and where finally $c_L$ is a short-hand notation to indicate the matrix element in the first row and first column of ${\cal S}(L)$, i.e., $c_L={\cal S}_{S,S}(L)$.
 The latter is the fundamental parameter of the model as it encodes
 %The above expressions shows that
 the functional dependence upon the ``temporal"  index $L$.
 %in the model is encoded in
%the matrix element $c_L$.
%plays a key role in the stroboscopic evolution of the system it
%encodes the functional
As detailed in the Appendix,  a closed  expression for $c_L$  is obtained by explicitly solving the scattering process, yielding
\begin{equation}\label{ELE}
c_L=\frac{\left(\lambda_+^L-\lambda_-^L\right)r_1+\lambda_+\lambda_-^L-\lambda_+^L\lambda_-}{\lambda_+-\lambda_-},
\end{equation}
with
\begin{eqnarray}
\lambda_{\pm}=\frac{1}{2}\left[\left(r_1e^{i\phi}-r_1r_2\right)\pm\sqrt{\left(r_1e^{i\phi}-r_1r_2\right)^2+4r_2e^{i\phi}}\right].
\end{eqnarray}

Equation~(\ref{EQ0}) describes a thermal bosonic Gaussian channel $ {\cal E}_{\eta}^{N}$ \cite{PRA}  associated with a thermal bath with mean photon number $n_T$ and with effective transmissivity
$\eta=|c_L|^2$,  i.e.,
\begin{eqnarray}\label{Emap}
\rho_S^{\inp} \longmapsto \rho_S^{\out,L} = {\cal E}_L ( \rho_S^{\inp} ) :=  {\cal E}_{|c_L|^2}^{n_T} ( \rho_S^{\inp} )\;.
\end{eqnarray}
For example, if the input state of $S$ is a coherent state $\rho^{\inp}_S=|\alpha\rangle\langle \alpha|$, Eq.~(\ref{EQ0})
predicts that the output state will be a
%one gets
%\begin{equation}
%\chi^{\out}_{|\alpha\rangle\langle\alpha|}(\nu_S)=\exp{\left[(n_T+\frac{1}{2}) (|c_L|^2-1)
%|\nu_S|^2+c_L^*\nu_S\alpha^*-c_L\nu_S^*\alpha\right]},
%\end{equation}
%which describes a
displaced thermal state. In particular if the environmental modes are initialized in the vacuum  (i.e. $n_T=0$), the output state
$\rho_S^{\out,L}$ is still a coherent state $|c^*_L\alpha\rangle$. We also notice that when the temperature of the environment is zero (i.e. $n_T=0$ again) and the input state of the mode $S$ spans  over the vector  subspace with at most one photon, the channel~(\ref{Emap})
can be described as a qubit amplitude damping channel
%It can be formally identified with a qubit amplitude damping channel only on the system qubit
${\cal A}_{\left|c_{L}\right|^2}$~\cite{NIELS} characterized by
%a transmissivity
%$\eta=\left|c_{L}\right|^2$, i.e.
%\begin{eqnarray}
%\rho_S^{\inp} \longmapsto \rho_S^{\out} = {\cal E}_{|c_{L}|^2} [ \rho_S^{\inp}]\;.
%\end{eqnarray}
%with
the following Kraus operators,
%\cite{NIELS},
\begin{equation}\label{EQ1}
E_0=\left(
      \begin{array}{cc}
        1 & 0 \\
        0 & c_{L} \\
      \end{array}
    \right), \quad
    E_1=\left(
          \begin{array}{cc}
            0 & \sqrt{1-|c_{L}|^2} \\
            0 & 0 \\
          \end{array}
        \right).
\end{equation}

%The above expressions shows that the functional dependence upon the ``temporal"  index $L$ in the model is encoded in
%the matrix element $c_L$.
%%plays a key role in the stroboscopic evolution of the system it
%%encodes the functional
%A closed  expression for such parameter is obtained by explicit solving the scattering process yielding, see Appendix for details,
%\begin{equation}\label{ELE}
%c_L=\frac{\left(\lambda_+^L-\lambda_-^L\right)r_1+\lambda_+\lambda_-^L-\lambda_+^L\lambda_-}{\lambda_+-\lambda_-},
%\end{equation}
%with
%\begin{eqnarray}
%\lambda_{\pm}=\frac{1}{2}\left[\left(r_1e^{i\phi}-r_1r_2\right)\pm\sqrt{\left(r_1e^{i\phi}-r_1r_2\right)^2+4r_2e^{i\phi}}\right].
%\end{eqnarray}
% are the eigenvalues of the reduced scattering matrix for $a_S$ and $a_0$, Eq. (\ref{Sr}), in the case of vacuum environments.

\section{Characterization for non-Markovianity}\label{characterization}
\subsection{Sufficient and necessary condition of non-Markovianity}
In this section, we will show a sufficient and necessary condition of the non-Markovianity of our channel by adopting the
divisibility condition analyzed in Refs.~\cite{RIVAS,WOLF,SAB}.
The idea is as follows.
The  map $\mathcal{E}_L$ which describes the evolved state of $S$ after $L$ collisional  steps, see Eq.~(\ref{Emap}), can be formally split as
\begin{equation}
\mathcal{E}_{L}=\Phi_{L-1\rightarrow L}\circ\mathcal{E}_{L-1},
\end{equation}
where $\Phi_{L-1\rightarrow L}$ is an intermediate process that maps the output state after the $(L-1)$-th step into the final state $\rho_S^{\out,L}$
and where ``$\circ$" represents the composition of super-operators. By construction both %From Eq. (\ref{Emap}) we see that
$\mathcal{E}_{L}$ and $\mathcal{E}_{L-1}$ are complete positive (CP) and trace preserving~\cite{NIELS}, i.e., they describe a proper
dynamical evolution.
 %\jj{[shall we add an explanation of why it is CPT?]}.
 On the other hand, there is no guarantee in general, that the complete positivity condition  should  hold for  $\Phi_{L-1\rightarrow L}$.
As discussed in Refs.~\cite{RIVAS,WOLF,SAB} this fact defines the divisibility property of the dynamical evolution of the system
which is an essential pre-requisite for Markovianity.
In particular if $\Phi_{L-1\rightarrow L}$ is non-CP for some value of $L$ then we say
that the process is {\it nondivisible} and hence non-Markovian. Vice-versa if $\Phi_{L-1\rightarrow L}$ is CP for all $L$ then the dynamics
is {\it divisible} and hence Markovian (at least in a generalized sense).
%
% Therefore if there exist steps $(L-1)$ and $L$ that $\Phi_{L-1\rightarrow L}$ is not CPT, the dynamics will be non-Markovian \cite{RIVAS}.

In our case an explicit expression for $\Phi_{L-1\rightarrow L}$ can be obtained by using Eq.~(\ref{EQ0}) to
show that, for arbitrary $L$, the inverse of the channel
 $\mathcal{E}_L$  is given by
 \begin{eqnarray}
\chi^{\out,L}(\nu_S)&\mapsto&\chi^{\inp}(\nu_S)=\mathcal {E}_L^{-1}[\chi^{\out}(\nu_S)]\\ &=
&\chi^{\out,L}\left(\frac{\nu_S}{c_{L}^*}\right)\exp{\left[\left(n_T+\frac{1}{2}\right)(1-|c_{L}|^2)\frac{|\nu_S|^2}{|c_{L}|^2}\right]}.\nonumber
\end{eqnarray}
Accordingly  we can write
\begin{eqnarray}\label{phim}
&&\Phi_{L-1\rightarrow L}[\chi^{\inp}(\nu_S)]=\mathcal{E}_L\circ\mathcal{E}_{L-1}^{-1}[\chi^{\inp}(\nu_S)]\cr\cr &=&\chi^{\inp}\left(c_r^*\nu_S\right)\exp{\left[\left(n_T+\frac{1}{2}\right)\left(\left|c_r\right|^2-1\right)|\nu_S|^2\right]},
\end{eqnarray}
where $c_r=c_L/c_{L-1}$.

In order to check whether or not $\Phi_{L-1\rightarrow L}$ is completely positive and trace preserving, we adopt the criterion proposed in Ref. \cite{Caruso}. By performing the transformation,
\begin{eqnarray}
\nu_S&=&\frac{1}{\sqrt{2}}(-y-ix),\cr\cr
\nu_S^*&=&\frac{1}{\sqrt{2}}(-y+ix),\cr\cr
|\nu_S|^2&=&\frac{1}{2}(x^2+y^2),
\end{eqnarray}
we can rewrite Eq. (\ref{phim}) as the following,
\begin{equation}
\Phi_{L-1\rightarrow L}[\chi^{\inp}(z)]=\chi^{\inp}\left(K\cdot z\right)\exp{\left(-\frac{1}{2}z^T\cdot\alpha\cdot z\right)},
\end{equation}
where $z=[x,y]^T$ is a column vector of $R^2$ and
\begin{eqnarray}
K&=&\left[
    \begin{array}{cc}
     \mathrm{Re}(c_r) & \mathrm{Im}(c_r) \\
     -\mathrm{Im}(c_r)  & \mathrm{Re}(c_r)\\
    \end{array}
  \right], \nonumber \\
    \alpha&=&\left(n_T+\frac{1}{2}\right)\left(1-\left|c_r\right|^2\right)
    \left[\begin{array}{cc}
    1 & 0 \\
     0  & 1\\
    \end{array}\right],
\end{eqnarray}
are $2\times2$ real matrices.

If the matrix defined as $M=2\alpha-\sigma_y+K^T\sigma_y K$ is non-negative definite then $\Phi_{L-1\rightarrow L}$ is CP and thus the channel is Markovian. It is easy to compute the eigenvalues of $M$ as follows,
\begin{eqnarray}
\lambda_1&=&2(n_T+1)\left(1-\left|c_r\right|^2\right),\cr\cr
\lambda_2&=&2n_T\left(1-\left|c_r\right|^2\right).
\end{eqnarray}
Thus we can conclude that the process is Markovian if and only if $|c_r|\le1$, i.e.,
\begin{equation}\label{inequa}
|c_{L}|\le|c_{L-1}|, \forall L\ge1, \quad \Longleftrightarrow \quad \mbox{Markovianity.}
\end{equation}
%the channel is Markovian.
We note that this criterion is independent of the environmental temperature $n_T$ and of the input state.

\subsection{Characterization of non-Markovianity of the channel}
\begin{figure}
  \includegraphics[width = 1\linewidth]{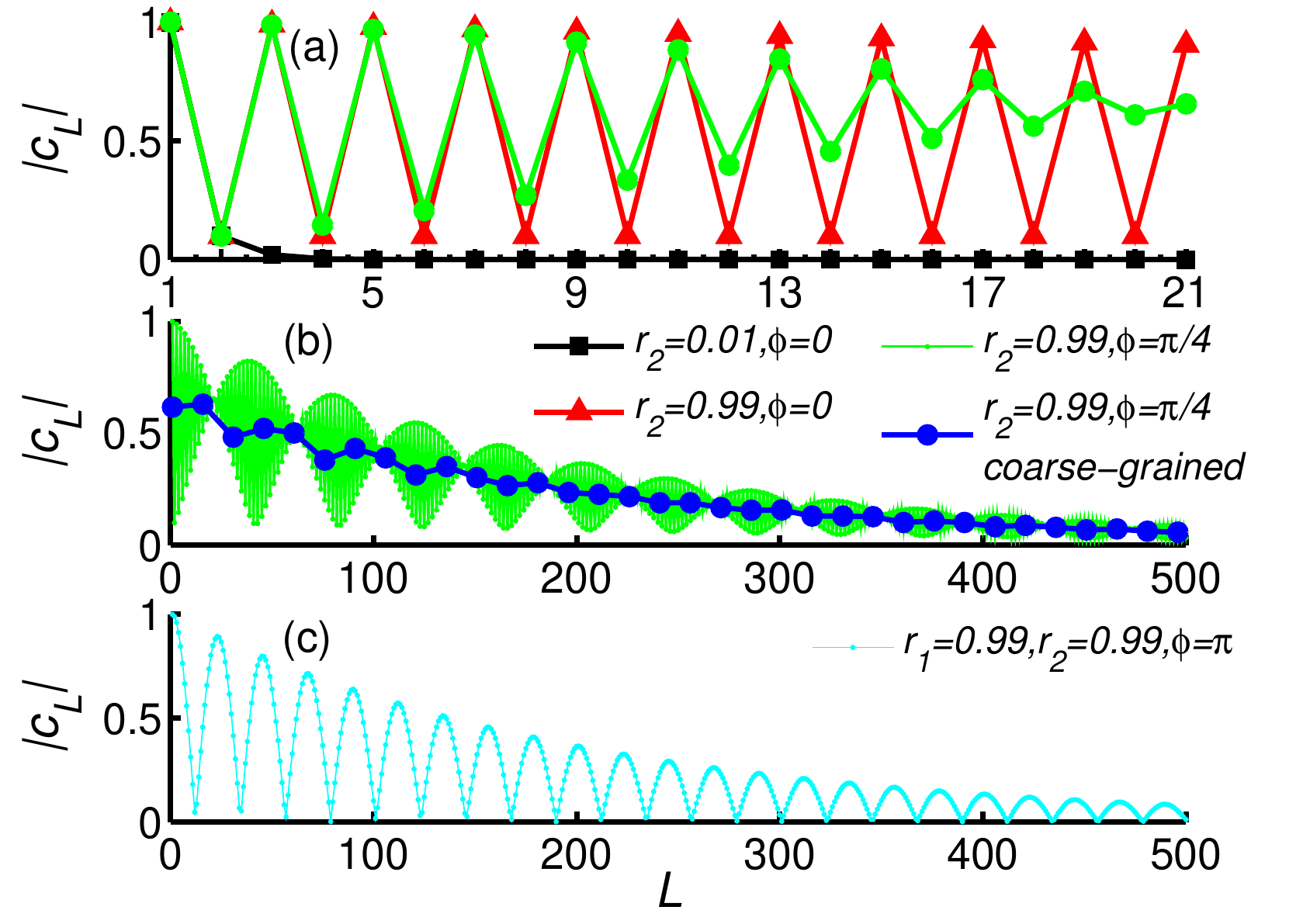}
  \caption{(Color online). (a) Stroboscopical and (b) coarse-grained evolutions of $|c_L|$ for $r_1=0.1$ and different values of the phase
  shift~$\phi$. For $r_2=0.01$ (black line, weak memory), $|c_L|$ decreases monotonically. For $r_2=0.99$ (red and green lines, strong memory), there is an evident oscillation of $|c_L|$. A nonzero phase (green lines) will modify the amplitude of the oscillation. The coarse-grained evolution (blue line) with grain size $\Delta=15$ also reveals the non-Markovian character of the channel. (c) Stroboscopically evolution of $c_L$ for $\phi=\pi$ in the limit of $r_1,r_2\rightarrow 1$. In this limiting case the stroboscopic evolution exhibits a kind of evolution under continuous limit.}
  \label{fig1a}
\end{figure}

In this section we adopt the criterion Eq. (\ref{inequa}) to characterize the divisibility of the process.
In Fig. \ref{fig1a} we show the stroboscopic evolution of $|c_L|$ with different reflectivities of \mbox{BS1} and BS2 and
for different values of the phase shift $\phi$. For a small $r_2$, which means a weak feedback from the environment, the monotonic decrease of $|c_L|$ implies the Markovianity of the channel. For a large $r_2$ instead the back-flow of information from the environmental modes into the system mode is strong: in this case  oscillations of $|c_L|$ are evident, testifying the non-Markovian character of the evolution.

It is interesting to investigate the coarse-grained evolution of $|c_L|$. We define the averaged value of $|c_L|$ in the following,
\begin{equation}
\left|\overline{c}_n\right|=\frac{1}{\Delta}\sum_{k=(n-1)\Delta}^{n\Delta}{|c_k|},
\end{equation}
where $n=1,2,...,L/\Delta$ and $\Delta$ is the size of the grain. In Fig. \ref{fig1a}(b) we show the coarse-grained evolution. The non-Markovianity of the channel still holds because of the oscillations of $|c_n|$. The existence of non-Markovianity in the coarse-grained evolution reveals the possibility of a survival of non-Markovianity of our channel in the continuous limit. In fact, we can obtain a kind of continuous limit evolution in the case of $\phi=\pi$ and $r_1,r_2\rightarrow 1$ without doing the coarse-grained computation, see Fig. \ref{fig1a}(c).

\begin{figure}
  \includegraphics[width = 1\linewidth]{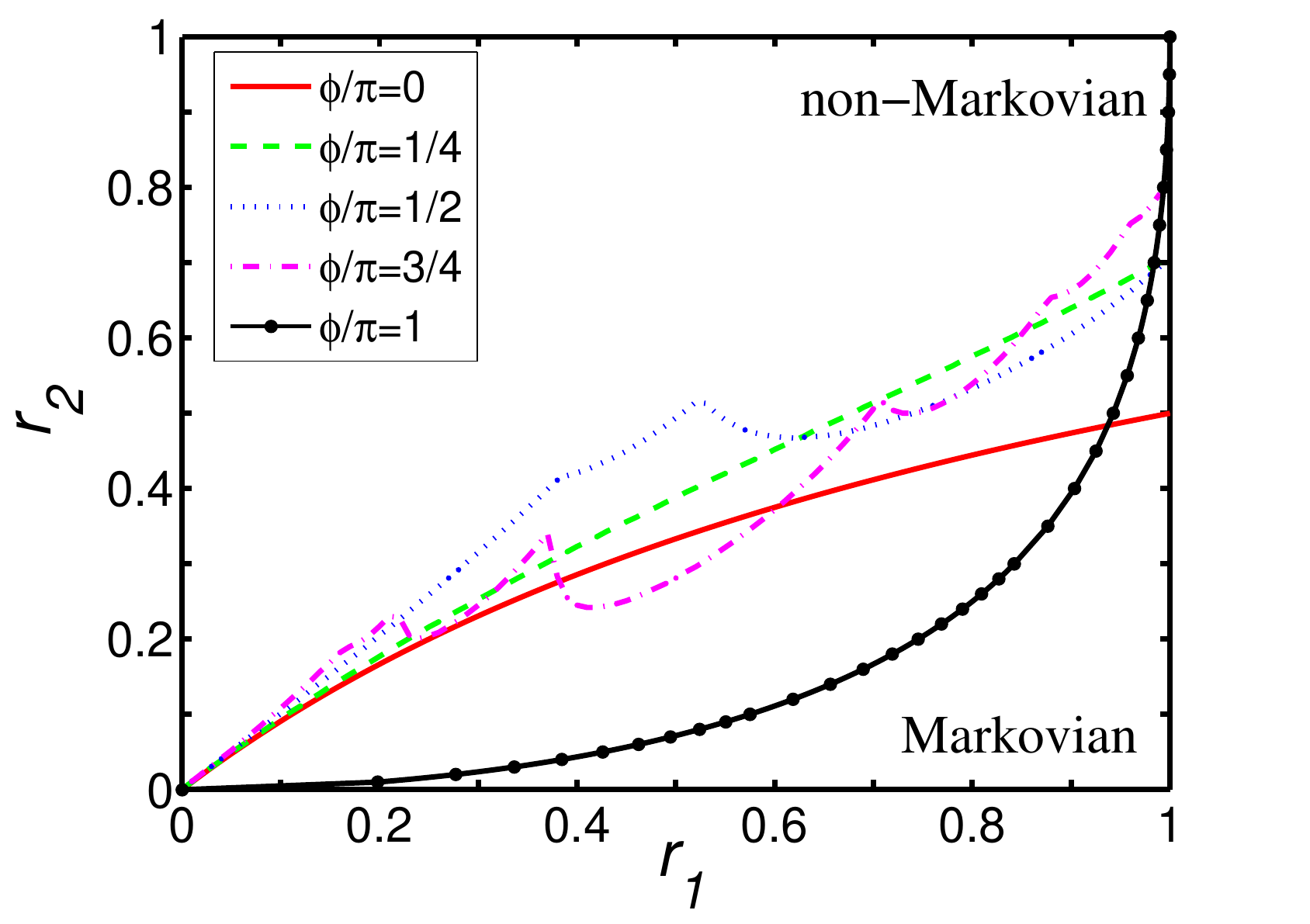}
  \caption{(Color online). Boundaries between non-Markovian and Markovian regions for different $\phi\in[0,\pi]$ in $r_1-r_2$ space. Each curve, corresponding to a fixed $\phi$, separates the space into two parts; the upper part is the non-Markovian region and the lower part is the Markovian region. The analytical expressions for the boundary $\phi=0$ is $r_2=2r_1/(1+r_1)$ and for $\phi=\pi$ it is $r_1=2\sqrt{r_2}/(1+r_2)$.}
  \label{fig2}
\end{figure}

\begin{figure}
  \includegraphics[width = 1\linewidth]{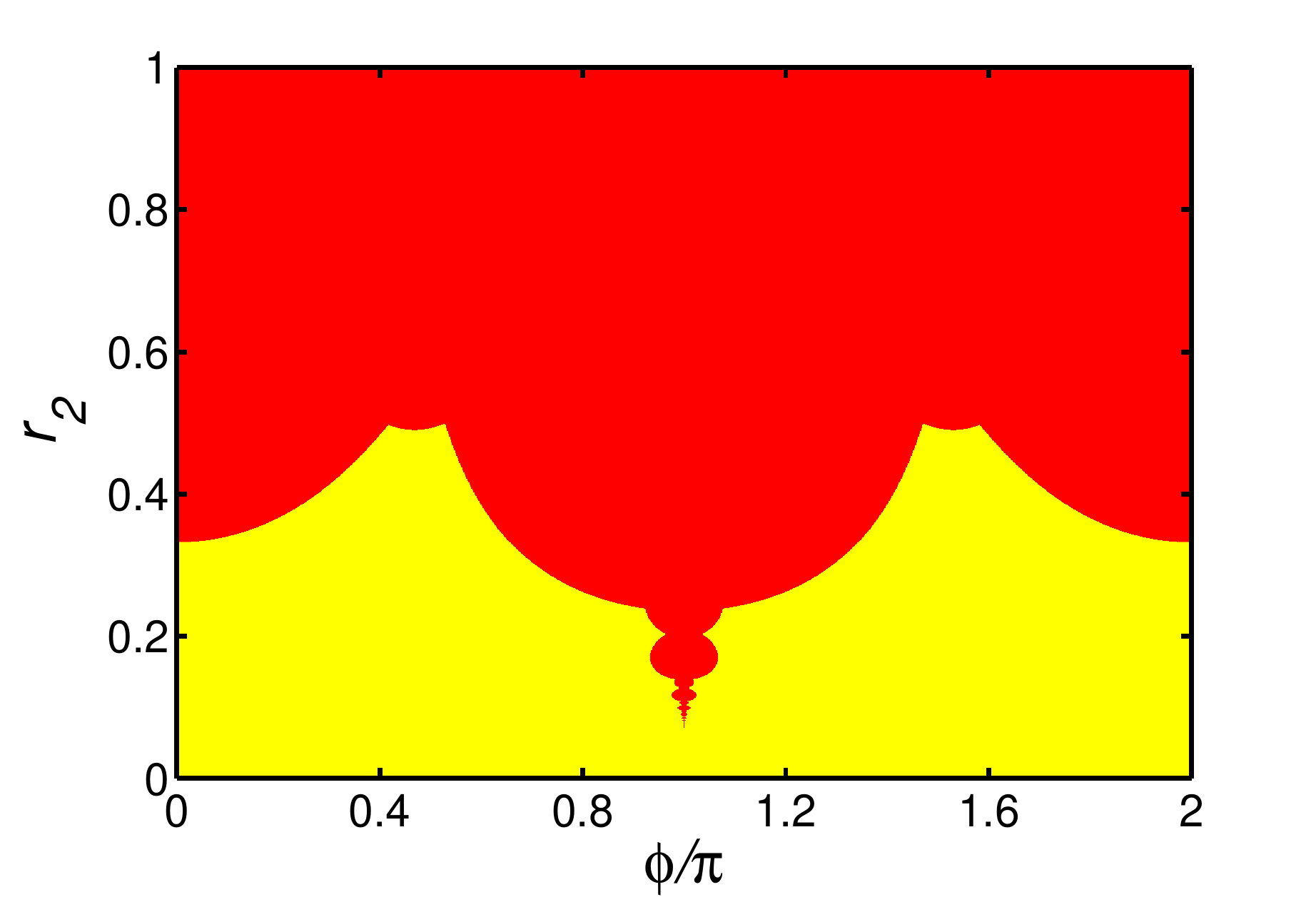}
  \caption{(Color online). Dependence of non-Markovianity on $\phi$. The reflectivity of BS1 is $r_1=0.5$. The red is the non-Markovian regime and the yellow is the Markovian regime. }
  \label{fig3}
\end{figure}

In Fig. \ref{fig2} we show the boundaries between the non-Markovian and Markovian regions for different $\phi\in[0,\pi]$ in $r_1-r_2$ space [the boundary for $\phi\in(\pi,2\pi]$ is the same as that of the $(2\pi-\phi)$ case]. We can see that for $r_2=0$ the channel is always Markovian regardless of the values of $r_1$ and $\phi$, because there is no back-flow information allowed. Although, in general, for a fixed $\phi$ the boundary is not a simple function of $r_1$ and $r_2$, we can obtain the analytical expressions for the boundaries for two specific cases $\phi=0$, and $\pi$. For $\phi=0$, the boundary is given by $r_2=2r_1/(1+r_1)$ and for $\phi=\pi$ the boundary is given by $r_1=2\sqrt{r_2}/(1+r_2)$.
%
%The boundary of a $\phi\in(\pi,2\pi]$ is the same as that of the $(2\pi-\phi)$ case.
In Fig. \ref{fig3} we show the dependence of non-Markovianity on $\phi$ in $\phi-r_2$ space. We see that the (non-)Markovian region is symmetric about $\phi=\pi$.

\section{Witnessing non-Markovianity of the channel} \label{NONM}

Several witnesses of non-Markovianity have been developed in the recent years. Even though they are typically identified under the name of ``measures" of non-Markovianity, they do not gauge the strength of the non-Markovian character of a dynamical evolution in an operational sense. Still such witnesses are useful theoretical tools that provide sufficient conditions that  can be used to certify the presence  of non-Markovianity in the physics of the problem.
%To be precise
%these are not really proper
% measures in the sense that they cannot gauge the strength of non-Markovianinty in an operational sense.  Instead they should be seen as witnesses which one can use in order to certify that the dynamical evolution of the system do exhibit
% some sort of non-Markovian behavior.
In the following we study few of them: specifically the entanglement criterion introduced by Rivas {\it et al.}~\cite{RIVAS}, the trace-distance criterion introduced by Breuer {\it et al.}~\cite{BR,BR1}, and the relative entropy criterion introduced by Chru\'{s}ci\'{n}ski and Kossakowski~\cite{rS1,rS2}. These criteria, except for the qubit case mapping discussed in Sec.~\ref{SUBSUB},  are analyzed by fixing $\phi=0$ for simplicity.
 %\subsection{Witnessing non-Markoviantity with CV inputs} \label{CVIN}
\subsection{Witnessing non-Markovianity with entanglement} \label{TMSV}

In this section we employ the idea proposed in Ref.~\cite{RIVAS} to check the non-Markovianity of our stroboscopic evolution. Accordingly the system $S$ is initially prepared in an entangled state with an isolated ancilla $A$, and then subject to the channel: if the entanglement is not strictly monotonically decreasing during the evolution then we can say that the channel is non-Markovian, the opposite being not necessarily true, i.e., having entanglement which is monotonically decreasing does not necessarily correspond to having a Markovian evolution.

Due to the Gaussian character of the mapping Eq. (\ref{Emap}) we find it convenient to consider
 a two-mode squeezed vacuum (TMSV) $|\mathrm{TMSV}(\xi)\rangle_{SA}$ as input entangled probing state (see Fig.~{\ref{fig1}).
We recall that the vector $|\mathrm{TMSV}(\xi)\rangle_{SA}$ can be expressed as $S(\xi)|0_a0_b\rangle$
where $S(\xi)$ is the two-mode squeezed operator
\begin{equation}
S(\xi)=\exp{\left(\frac{1}{2}\xi^* ab-\frac{1}{2}\xi a^\dagger b^\dagger\right)}\;,
\end{equation}
with $a$ and $b$ being  the annihilation operators of the $S$ and $A$ modes respectively, and with $\xi=re^{i\theta}$ being the squeezing parameter (without loss of generality  we set $\theta=0$).
%Assume then that the environmental ancillary mode of the scheme in Fig.~\ref{fig1} are all initialized into thermal states having the same temperature. The joint input state of the system $S$, $A$ plus environment is hence given by
%\begin{equation}
%\rho^{\mathrm{IN}}=|TMSV(\xi)\rangle_{AS}\langle TMSV(\xi)|\otimes\rho^{\mathrm{th}}_1\otimes\rho^{\mathrm{th}}_2\otimes...\otimes\rho^{\mathrm{th}}_N
%\end{equation}
%where $\rho^{\mathrm{th}}_i=\frac{1}{1+n_T}\left(\frac{n_T}{1+n_T}\right)^{a_i^\dagger a_i}$, $n_T$ being the thermal photon number.
Following the derivation of the previous section we can then express the output characteristic function of the $A$ and $S$ as
%\begin{widetext}
\begin{eqnarray}
&&\chi^{\mathrm{out},L}_{AS}(\nu_A,\nu_S)
%=\chi^{\mathrm{out},L}_{AS}(\overrightarrow{\nu})=\chi^{\mathrm{in}}_{AS}(\mathcal{S}^{-1}\overrightarrow{\nu})\cr\cr
=\exp\big[-\frac{|\nu_A|^2+|c_{L}|^2|\nu_S|^2}{2}\cosh{\xi} \\&&-\frac{c_{L}^*\nu_A\nu_S+c_{L}\nu_A^*\nu_S^*}{2}\sinh{\xi}
-\frac{(2n_T+1)(1-|c_{L}|^2)|\nu_S|^2}{2}\big]. \nonumber
\end{eqnarray}
%\end{widetext}
%where $\overrightarrow{\nu}=[\nu_A,\nu_S,\nu_1,\nu_2,...,\nu_N]$ and $c_{L}$ defined as before.
This is a Gaussian state with  covariance matrix
\begin{equation}
V=\frac{1}{2}\left(
               \begin{array}{cc}
                 B_1 & B_3 \\
                 B_3^T & B_2 \\
               \end{array}
             \right),
\end{equation}
where $B_1=\cosh{\xi}I_2$, $B_2=[(1-|c_{L}|^2)(2n_T+1)+|c_{L}|^2\cosh{\xi}]I_2$, and $B_3=-c_L\sinh{\xi}\sigma^z$ ($\sigma^z$ is the Pauli matrix). We notice that the Gaussianity of the output system state is preserved after tracing out the ancilla mode.
 %For vacuum environment, the result can be obtained by just simply setting $n_T=0$.
Its entanglement can hence be faithfully measured by the logarithmic negativity~\cite{LOG}
\begin{equation}
E_N=\max{\{-\ln{2\mu},0\}}, \label{ENTA}
\end{equation}
\begin{equation}
\mu=\sqrt{\frac{\Sigma-\sqrt{\Sigma^2-4\det{[V]}}}{2}},
\end{equation}
where $\Sigma=\det{[B_1/2]}+\det{[B_2/2]}-2\det{[B_3/2]}$.

We note that the squeezing parameter $\xi$, which parameterizes the entanglement of the testing TMSV state, does not affect the existence of the oscillation of $|c_L|$ during the stroboscopical evolution; it affects only the amplitude of the oscillation. A larger value of $\xi$ means a more entangled TMSV state and will lead to a more pronounced oscillation, if it exists. Therefore we can use in principle any nonzero value of $\xi$ to witness non-Markovianity. In Fig.~\ref{fig5} the entanglement criterion is tested for the case of thermal environments ($n_T\geq0$) with different $r_1$,
by looking at the functional dependence of~(\ref{ENTA})
with respect to the temporal parameter $L$.
We see that as the environment temperature increases above a threshold the region where the criterion based on entanglement revival can be used to certify the presence of non-Markovianity shrinks (the thermal noise being too strong to maintain entanglement in the system). The dependence of the threshold temperature on $r_1$ is $n^c_{T}=r_1^2/(1-r_1^2)$, as shown in the inset of Fig.~\ref{fig5}. For $r_1=1$ the threshold temperature is infinite because the system completely decouples from the environments. For $r_1=0$ the threshold temperature is zero, which means that the criterion is valid only in the case of a vacuum environment.
\begin{figure}
  \includegraphics[width = 1\linewidth]{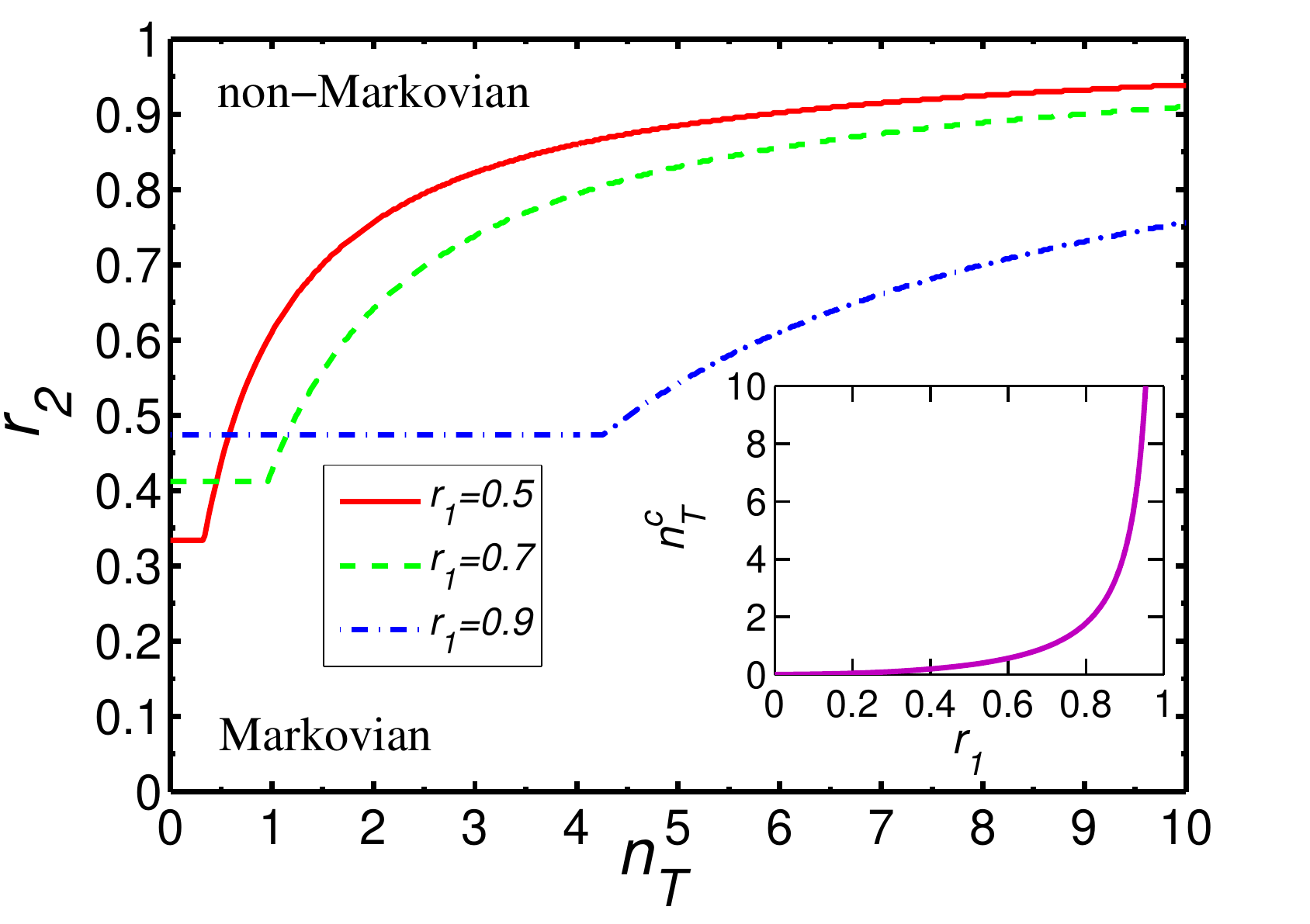}
  \caption{(Color online). The non-Markovian region detected by the criterion based on entanglement revival \cite{RIVAS} in the $n_T-r_2$ plane. Each curve, corresponding to $r_1=0.5, 0.7$ and $0.9$, separates the plane into two parts. The squeezing parameter of the testing TMSV state is $\xi=1$, although a different value of $\xi$ does not affect the results. The non-Markovian region is above this curve: for this points out in fact that the quantity Eq. (\ref{ENTA}) is not monotonically decreasing with $L$.
  Below the curve the method instead fails to identify the non-Markovian character of the process (the quantity Eq.~(\ref{ENTA}) being monotonic in $L$). The flat part of each curve is exactly the same as the boundary predicted by Eq. (\ref{inequa}). The inset shows the dependence of the threshold temperature $n^c_T$, above which the entanglement revival criterion is less valid to detect non-Markovianity, on the reflectivity~$r_1$.}
  \label{fig5}
\end{figure}

\subsubsection{Witnessing non-Markovianity of the qubit channel}\label{SUBSUB}
As discussed at the end of Sec.~\ref{scheme}, the channel ${\cal E}_L$ reduces to the qubit amplitude damping channel  ${\cal A}_{\left|c_{L}\right|^2}$ in the case of $n_T=0$ and when the input states of the model are restricted to superpositions of Fock states involving no more than a single photon.

The non-Markovianity of such mapping can also be studied
via the entanglement criterion proposed in Ref.~\cite{RIVAS}. In this case we can suppose that we have a maximally entangled state between the qubit input system and an external qubit ancilla (mode $A$), e.g., a state of the form  $|\psi\rangle_{SA}=(|0_S1_A\rangle+|1_S0_A\rangle)/\sqrt{2}$.
The evolved reduced density matrix of system and ancilla becomes hence
\begin{equation}
\rho^{\mathrm{out},L}_{AS}=\frac{1}{2}\left(
                         \begin{array}{cccc}
                           1-\left|c_{L}\right|^2 & 0 & 0 & 0 \\
                           0 & \left|c_{L}\right|^2 & c_{L}^* & 0 \\
                           0 & c_{L} & 1 & 0 \\
                           0 & 0 & 0 & 0 \\
                         \end{array}
                       \right),
\end{equation}
which is nothing but the Choi-Jamiolkoswki state of the transformation~\cite{REP}.
Its entanglement can then be measured by the concurrence \cite{WOOTTERS}. After a simple calculation one obtains
\begin{eqnarray}
E_C(\rho^{\mathrm{out},L}_{AS})=|c_{L}| \;.
\end{eqnarray}
The monotonicity of entanglement during the evolution therefore coincides with the monotonicity  of $|c_L|$. Accordingly, at variance with  the Gaussian channel analysis of the previous section, the cases under which the entanglement based criterion is able to detect the presence of non-Markovianity coincide with those associated with Eq. (\ref{inequa}).

The entanglement's stroboscopic evolution with different $\phi$ is shown in Fig. \ref{fig8}. We see that, for the given parameters, the channel , at first in a non-Markovian region, enters into a Markovian one ($\phi=\pi/2$) and then returns to a non-Markovian condition.
\begin{figure}
  \includegraphics[width = 1\linewidth]{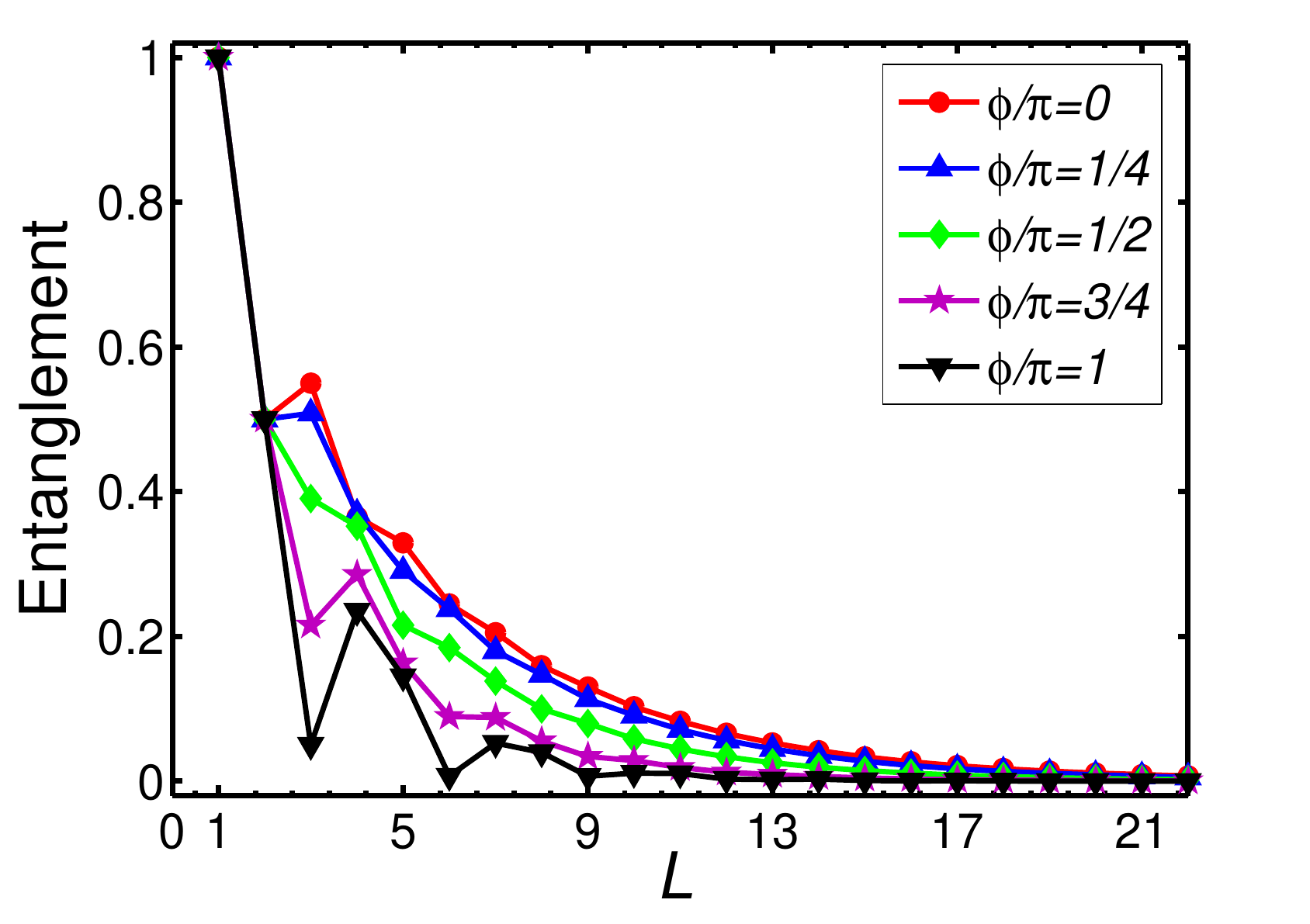}
  \caption{(Color online). Entanglement (concurrence) stroboscopic evolution with different $\phi$ for the qubit case. The parameters are chosen as $r_1=0.5$ and $r_2=0.4$.}
  \label{fig8}
\end{figure}

\subsection{Witnessing non-Markovianity with Gaussian fidelity} \label{CVfid}

In Ref. \cite{BR} the authors proposed to use the distance between two distinct input states to measure the non-Markovianity of a channel. In what follows we adopt a variant of this criterion proposed in Ref. \cite{BR1} which is well suited to study the evolution of Gaussian input states. The basic idea is the following: Consider two different states, say $\rho_1$ and $\rho_2$; if the channel is Markovian then the fidelity between these two states is monotonically increasing in the time evolution; if the monotonic increasing of fidelity is broken, then the channel is non-Markovian.

We recall that for two single-mode Gaussian states, $\rho_1$ and $\rho_2$, the fidelity can be calculated as follows~\cite{SCUTARU,Weedbrook},
\begin{equation}
F(\rho_1,\rho_2)=\frac{2}{\sqrt{\Delta+\delta}-\sqrt{\delta}}e^{-\frac{1}{2}\mathbf{\mathrm{d}}^T(\mathbf{V}_1+\mathbf{V}_2)^{-1}\mathbf{\mathrm{d}}},
\end{equation}
where $V_1$ and $V_2$ are the covariance matrix of $\rho_1$ and $\rho_2$, respectively. $\Delta=4\det{[V_1+V_2]}$, $\delta=(4\det{[V_1]}-1)(4\det{[V_2]}-1)$, and $\mathbf{\mathrm{d}}=\bar{x_2}-\bar{x_1}$ with $\bar{x_i}=[\langle q_i\rangle,\langle p_i\rangle]^T$.
Fig.~\ref{fig6} shows the results of the non-Marikovian detection with squeezed vacuum states. The squeezing parameters of the input states are $\xi_1=1$ and $\xi_2=0.5$ (other squeezing parameters will give the same result). Note that for $r_1=0$ the Markovian region appears only for $r_2=0$. Compared with Fig.~\ref{fig3} it can be seen that by using the fidelity we can detect larger non-Markovian regions, moreover, the boundary is independent of environment temperature $n_T$. The non-Markovian region in Fig. \ref{fig6} coincides with that predicted by Eq. (\ref{inequa}). We also note that, by choosing two coherent states as probes, the same results can be obtained.
\begin{figure}
  \includegraphics[width = 1\linewidth]{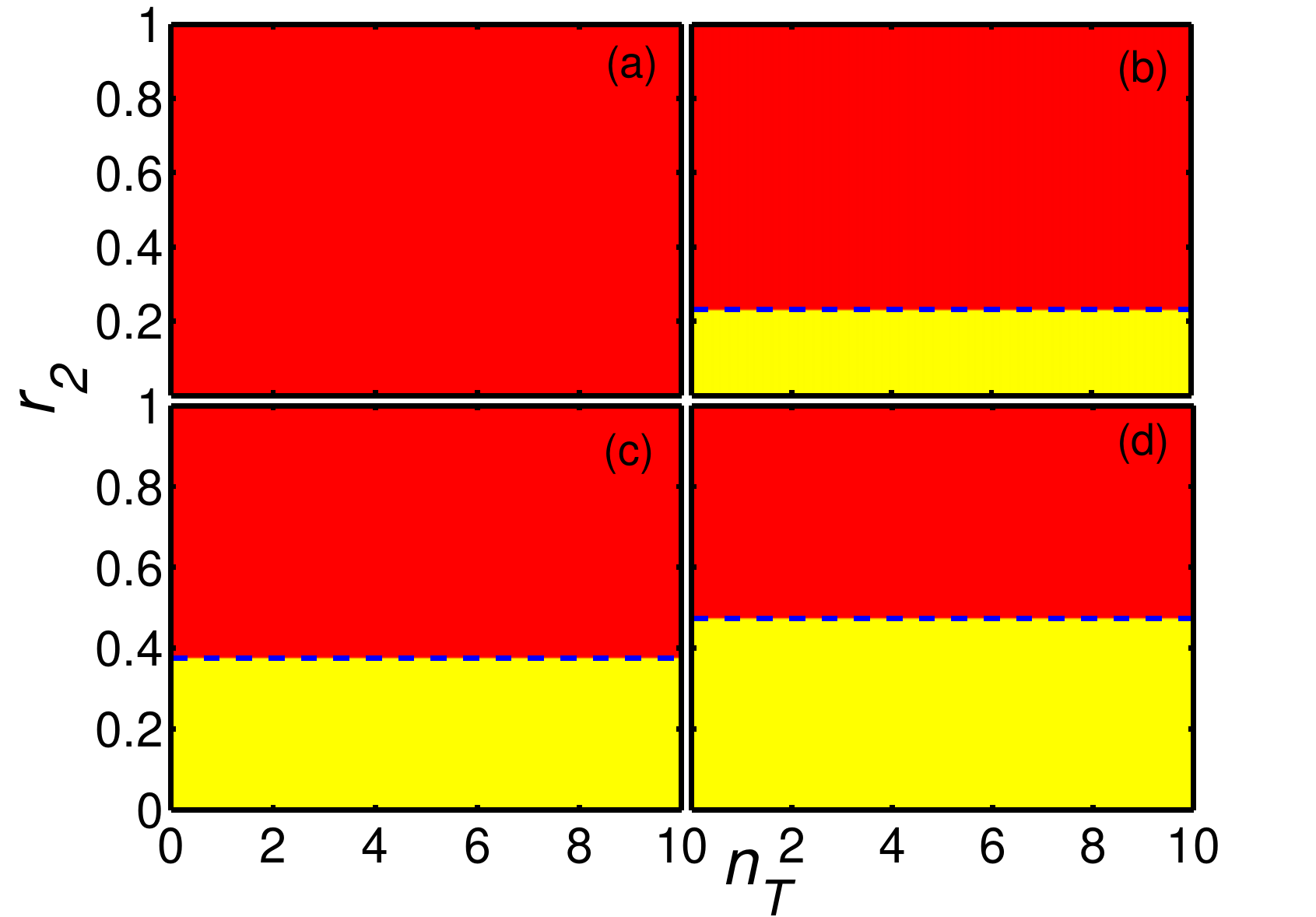}
  \caption{(Color online). Non-Markovian regions detected by the fidelity  \cite{BR,BR1} and relative entropy \cite{rS1,rS2} of two different squeezed vacuum states in the $n_T-r_2$ plane. The boundaries detected by these two criteria are exactly the same. The parameters are chosen as (a) $r_1=0$, (b) $r_1=0.3$, (c) $r_1=0.6$, (d) $r_1=0.9$, and $\phi=0$. The squeezing parameters of the two input states are $\xi_1=1$ and $\xi_2=0.5$, respectively. The red and yellow regions are non-Markovian and Markovian regions, respectively. The non-Markovian region coincides with that predicted by Eq. (\ref{inequa}) (dashed line).}
  \label{fig6}
\end{figure}

\subsection{Witnessing non-Markovianity with relative entropy}
As a final criterion we used the relative entropy criterion of Refs.~\cite{rS1,rS2}. Although the relative entropy is not a metric, it is acceptable as a measure of distinguishability.
The expression of relative entropy for a single mode Gaussian state $\rho_1$ with respect to another state $\rho_2$ is given by \cite{MAR},
\begin{equation}
S(\rho_1||\rho_2)=\mathrm{Tr}\left(\rho_2\ln{\rho_2}-\rho_2\ln{\rho_1}\right).
\end{equation}

Since the non-Markovianity of the channel allows the back-flow of information, if there is a revival of the relative entropy of two input Gaussian states during the evolution, the channel is non-Markovian. The non-Markovian regions detected by monotonicity of the relative entropy of two squeezed vacuum states during the evolution are shown in Fig. \ref{fig6}, and are exactly the same as those detected by the criterion based on fidelity. As a consequence, they also coincide with those predicted by Eq. (\ref{inequa}).

\section{Conclusions and outlook}
In conclusion, we proposed an all-optical scheme to simulate non-Markovian dynamics based on the collisional model. The interactions of system and environmental modes and the back flow of the information are implemented via BSs. We find that this channel is equal to a thermal bosonic channel and, if the environments are vacuum, it is reduced to the qubit amplitude damping channel. By properly tuning the reflectivities and the phase differences, we can switch our channel from the Markovian to the non-Markovian case.

We proved a sufficient and necessary condition for the non-Markovianity of the channel based on Gaussian inputs. We also investigated a few criteria that can be used to detect non-Markovian behaviors. We find that the criteria based on trace distance and relative entropy are stronger than the criterion based on entanglement revival in detecting non-Markovianity. The latter is, in fact, fragile when the environment is at high temperature.

Finally, we would like to comment on the possible experimental realization of this quantum simulator, which is based on concatenated beam splitters of tailored reflectivity and phase shifters (see Fig. \ref{fig1}). This device could be implemented in integrated quantum photonics on different technological platforms \cite{Politi,Marshall,Schreiber}, with the benefit of high stability, arbitrary control of the device parameters and improved scalability. In the proposed stroboscopic model the simulation of different Markovian and non-Markovian dynamics relies on the precise control of reflectivities in beam splitters and delays in phase-shifters. Such technological capability has already been demonstrated in several integrated photonic experiments \cite{Schreiber,Shadbolt,Crespi1,Tillmann,Crespi2}, both in static and dynamic approaches.

\acknowledgments
J. J. acknowledges discussions with F. Ciccarello. This work was supported by PRIN-MIUR (Project No. 2010LLKJBX), by EU (projects IP-SIQS and STREP-TERMIQ),
and by the National Natural Science Foundation of China under Grants No. 11175033 and 11305021.

\begin{appendix}
\section{Calculation of $c_L$}
In this appendix we will show the derivation of the explicit expression for $c_L$. We use $a_S$ and $a_j$ ($j=0,1,...,L$) to denote annihilation operators of the input system and environment modes, while $a_{S'}$ and $a_{j'}$ ($j=0,1,...,L$) to denote annihilation operators of the corresponding output modes. The channel with $L$ steps is characterized by the scattering matrix ${\cal S}(L)$.

Now let us study of the physics of the element $c_L$ [i.e.,${\cal S}_{S,S}(L)$]. Note that after passing through the channel, the annihilation operators undergo the following transformation,
\begin{equation}
{\cal S}(L)\overrightarrow{a}^{\inp}=\overrightarrow{a}^{\out},
\end{equation}
with $\overrightarrow{a}^{\inp}=[a_S,a_0,a_1,...,a_L]^T$ being a vector of the input modes and $\overrightarrow{a}^{\out}=[a_{S'},a_{0'},a_{1'},...,a_{L'}]^T$ being a vector of the output modes. Therefore, the annihilation operator of the output mode can be expressed as a combination of the input modes as follows,
\begin{equation}
a_{j'}=\sum_{k=S}^{L}{{\cal S}_{j,k}(L)a_{k}},
\end{equation}
and the inverse transformation, in terms of the creation operators, from output to input modes,
\begin{equation}\label{invcreate}
a^\dagger_{j}=\sum_{k=S}^{L}{{\cal S}_{k,j}(L)a^\dagger_{k'}}.
\end{equation}
From Eq. (\ref{invcreate}) we see that the physics of the modulus of ${\cal S}_{k,j}(L)$ is the contribution of the $j$th input to the $k'$th output with other inputs being vacuum. In particular, for the case of $j=S$, $c_L$ [i.e. ${\cal S}_{S,S}(L)$] is the contribution of the system input to the system output with vacuum environments.

After understanding the physics for $c_L$, we can use a simplified model to derive the expression of $c_L$. In the simplified model, we consider the modes $0,1,...L$ to be vacuum. The reduced scattering matrix for modes $a_S$ and $a_0$ after $L$ steps is given by $\widetilde{{\cal S}}(L)=\widetilde{{\cal S}}_r^L$, where $\widetilde{{\cal S}}_r$ is the following $2\times2$ matrix,
\begin{equation}\label{Sr}
\widetilde{{\cal S}}_r=\left(
                         \begin{array}{cc}
                           r_1e^{i\phi} & \sqrt{1-r_1^2}e^{i\phi} \\
                           \sqrt{1-r_1^2}r_2 & -r_1r_2 \\
                         \end{array}
                       \right).
\end{equation}
According to the Jordan decomposition \cite{Horn}, we can express $\widetilde{{\cal S}}(L)$ as $\widetilde{{\cal S}}(L)=UVU^{-1}$, where $U=[\psi_+,\psi_-]$ and $V=\mathrm{diag}[\lambda^L_+,\lambda^L_-]$ are $2\times2$ matrices with $\psi_{\pm}$ being the eigenvectors of $\widetilde{{\cal S}}_r$ and $\lambda_{\pm}$ being the corresponding eigenvalues. The results for $\psi_{\pm}$ and $\lambda_{\pm}$ are the following,
\begin{eqnarray}\label{eiglambda}
\psi_{\pm}&=&\left[\frac{\sqrt{1-r_1^2}e^{i\phi}}{\lambda_{\pm}-r_1e^{i\phi}},1\right]^T,\cr\cr
\lambda_{\pm}&=&\frac{1}{2}\left(r_1e^{i\phi}-r_1r_2\pm\sqrt{\left(r_1e^{i\phi}-r_1r_2\right)^2+4r_2e^{i\phi}}\right).
\end{eqnarray}

Thus we can obtain the explicit expression of $c_L$ as follows; it is equal to the element in the first row and first column of $\widetilde{{\cal S}}(L)$,
\begin{equation} \label{cL}
c_L=\widetilde{{\cal S}}_{S,S}(L)=\frac{\left(\lambda^L_+-\lambda^L_-\right)r_1e^{i\phi}+\lambda_+\lambda_-^L-\lambda^L_+\lambda_-}{\lambda_+-\lambda_-}.
\end{equation}
\end{appendix}

\end{document}